# AI-enabled Automatic Multimodal Fusion of Cone-Beam CT and Intraoral Scans for Intelligent 3D Tooth-Bone Reconstruction and Clinical Applications

Jin Hao[1,3#], Jiaxiang Liu[2,4#], Jin Li[4], Wei Pan[4], Ruizhe Chen[2,4], Huimin Xiong[2,4], Kaiwei Sun[4], Hangzheng Lin[4], Wanlu Liu[5], Wanghui Ding[2], Jianfei Yang[4], Haoji Hu[4], Yueling Zhang[1], Yang Feng[6], Zeyu Zhao[4], Huikai Wu[4], Youyi Zheng[7], Bing Fang[8], Zuozhu Liu[2,4*], Zhihe Zhao[1]*

[1] State Key Laboratory of Oral Diseases & National Clinical Research Center for Oral Diseases & West China Hospital of Stomatology, Sichuan University, Chengdu, China
[2] Stomatology Hospital, School of Stomatology, Zhejiang University School of Medicine
[3] Harvard School of Dental Medicine, Harvard University, Boston, MA, USA
[4] Zhejiang University-University of Illinois at Urbana-Champaign Institute, ZJU-Angelalign R&D Center for Intelligent Healthcare, Zhejiang University, Haining, China
[5] Zhejiang University-University of Edinburgh Institute, Zhejiang University, Haining, China
[6] AngelAlign Tech Inc., Shanghai, China
[7] College of Computer Science and Technology, Zhejiang University, Hangzhou, China
[8] Ninth People's Hospital Affiliated to Shanghai Jiao Tong University, Shanghai Research Institute of Stomatology, National Clinical Research Center of Stomatology, Shanghai, China

[#]Equal Contribution.
*Correspondence: zuozhuliu@intl.zju.edu.cn (Z.L.) & zhzhao@scu.edu.cn (Z.Z.)

## Abstract

Cone-beam computed tomography (CBCT) is one of the most widely used digital models in dental practices for virtual treatment planning and patient management. A critical step in virtual treatment planning is to accurately delineate all tooth-bone structures from CBCT with high fidelity and accurate anatomical information. Previous studies have established several methods for CBCT segmentation using deep learning. However, the inherent resolution discrepancy of CBCT and the loss of occlusal and dentition information largely limited its clinical applicability. Here, we present a Deep Dental Multimodal Analysis (DDMA) framework consisting of a CBCT segmentation model, an intraoral scan (IOS) segmentation model (the most accurate digital dental model), and a fusion model to generate 3D fused crown-root-bone structures with high fidelity and accurate occlusal and dentition information. Our model was trained with a large-scale dataset with 503 CBCT and 28,559 IOS meshes manually annotated by experienced human experts. For CBCT segmentation results, we use a five-fold cross validation test, each with 50 CBCT, and our framework achieves an average Dice coefficient and IoU of 93.99% and 88.68%, respectively, significantly outperforming the baselines. For IOS segmentations, our model achieves an mIoU of 93.07% and 95.70% on the maxillary and mandible on a test dataset of 200 IOS meshes, which are 1.77% and 3.52% higher than the state-of-art method. In addition, our method for model fusion leads to a success registration rate of 91%, much higher than the selected baselines (13.3% or 6.67%). Our DDMA framework takes about 20 to 25 minutes to generate the fused 3D mesh model following the sequential processing order, compared to over 5 hours by human experts. Notably, our framework has been incorporated into a software by a clear aligner manufacturer, and real-world clinical cases demonstrate that our model can visualize crown-root-bone structures during the entire orthodontic treatment and can predict risks like dehiscence and fenestration. These findings demonstrate the potential of multi-modal deep learning to improve the quality of digital dental models and help dentists make better clinical decisions.

**Introduction**

Digital technology provides the promise of changing every aspect of modern dentistry, from virtual treatment planning to remote patient management. Indeed, the utilization of digital cone-beam computed tomography (CBCT) models has accelerated treatment planning and management in dental practices, like in orthodontics and implant surgery. One advantage of these CBCT models is that they can provide complex anatomical structures of both teeth and bones, creating a virtual model of the head [1,2]. Moreover, accurate CBCT models can facilitate computer-aided implant planning, simulation of prosthetic evaluation, and prediction of orthodontic outcomes [3]. To achieve such goals, one fundamental component of this process is the accurate 3D segmentation of teeth, jaws, and bony structures from CBCT images. However, it is challenging and laborious to segment the teeth and bones from CBCT images due to the limited resolution, high noise levels, and cone beam artifacts [4], resulting in significant human labor and compromised quality of model visualization and segmentation outcomes.

Recent efforts have been made towards the automatic segmentation of teeth and bones in CBCT images based on deep learning [5-9]. These approaches can generate highly efficient segmentation results of tooth-bone structures for digital model visualization. However, the clinical applicability of such segmented CBCT images is still limited. Previous research has shown that CBCT measurements can be 6.9% smaller than the actual values of objects depending on the resolutions [10,11]. This CBCT value discrepancy, termed shrinkage, makes the digital model unreliable for precise simulation of implant surgery and orthodontic outcomes, suggesting the necessity of data matching between CBCT and other modality measurements in clinical practices. In addition, CBCT often failed to recapitulate accurate information of the occlusal surfaces due to the high density of enamel, dental restorations, implants, and orthodontic appliances [4,12,13]. Accurate capturing of occlusal surface information is of great significance in many clinical applications, as the design of saw guides, drilling guides, and orthognathic positioning guides are largely dependent on occlusal surfaces [14-16]. Furthermore, the correct anatomical position of the dentition in the maxilla and mandible is often missing in CBCT images, making orthodontic treatment simulation impossible based on CBCT. Therefore, accurate matching of the dentition information in CBCT segmentation is required for its application in broader clinical practices.

Compared to CBCT, the intraoral scanners (IOSs) are widely used in dentistry to replace impressions and study casts nowadays. They can generate a digital impression of the tooth's anatomy by projecting a light source on the dental arches. Although the IOS models only provide images of the crown and gingiva, the scans are considered accurate if not more accurate than plaster models. Previous research has shown that the shrinkage ratio of IOS models can be as low as 0.9%, the most accurate in 3D digital modalities [10]. In addition, the IOS are sufficiently accurate for capturing occlusal information and position of the dentition [17]. Therefore, the IOS models could provide the accuracy and occlusal information that is needed in CBCT for clinical applications. Recently, our group and others have developed methods for automatic segmentation of the IOS models using deep learning with high accuracy [18-21]. Herein, the automatic fusion of both IOS and CBCT segmentations could provide crown-root-bone structures with the accuracy needed in many dental applications.

Our study aims to reconstruct the crown-root-bone structures by fusing the IOS mesh and CBCT image data with deep learning. We build a multi-modal framework consisting of a CBCT segmentation model, an IOS segmentation model, and a fusion model to generate highly accurate crown-root-bone structures with high resolution. In addition, we demonstrate the clinical applicability of our model, named Deep Dental Multimodal Analysis (DDMA), with real-world clinical cases from malocclusion patients. By simulating the crown-root-bone structures during orthodontic tooth movement, we provide clinical evidence that DDMA can predict most of the risks during orthodontics, like fenestration and alveolar ridge resorption. In addition, DDMA can accurately reconstruct crown-root-bone relationships after orthodontic tooth movement using intraoral scans, with no additional CBCT scans needed.

In this study, we show that accurate simulation and treatment planning is scarcely possible to achieve with only CBCT scans with a rigorous study on 100 patients. The CBCT segmentation model achieves an average IoU and Dice coefficient of 88.68% and 93.99% on a five-fold cross-validation test with 503 patients. The IOS segmentation model achieves an mIoU and per-face accuracy of an mIoU of 93.07% and 95.70% on the maxillary and mandible on a hold-out set with 200 IOS scans, significantly outperforms the current state-of-the-arts. As for multimodal fusion, the generated fused mesh achieves mean Chamfer distance (CD) and average symmetric surface distance (ASSD) of 0.18mm and 0.20mm tested on a hold-out set of 50 patients. Finally, our DDMA framework is 2 to 4 magnitudes faster than human experts to generate CBCT and IOS segmentations, and 10x faster to generate fused meshes than human experts with assistance from interactive software. We demonstrate the clinical applicability of the DDMA framework with real-world cases from patients seeking for orthodontics treatment. The results reveal that the DDMA can help avoid fenestration, accurately simulate the treatment effect during the long-term treatment process by taking only one CBCT scan. In summary, our DDMA is the first clinical applicable multi-modal framework based on deep learning for accurate segmentation, simulation, and diagnostic analysis of the tooth crown-root-bone structures in digital dentistry.

**Results**

**Experimental Setup.** We conducted experiments on a large-scale dataset which contains 503 samples with both CBCT and IOS, and an extra 28,559 IOS meshes collected from hospitals and clinics in 25 provinces in China during 2018-2021. The 503 patients are at age 19.53±7.57 years old, with 32.5% male and 67.5% female. The CBCT images and IOS meshes were annotated by groups of human experts. For CBCT segmentation, each pixel was associated with a human experts' annotation for background and tooth or alveolar bone. For IOS segmentation, every mesh face fi was associated with a human experts' annotation yi, where $y_i \in \{0, 11-18, 21-28, 31-38, 41-48\}$ denotes the gingiva and FDI notation of 32 different teeth. Note that both CBCT and IOS data are acquired by different types of equipment, e.g., CBCT images are collected from equipment with resolution ranges from 0.125mm to 0.5mm. Detailed data statistics and experimental settings were reported in the Supplementary.

**Module Details of DDMA.** The DDMA framework contained three major modules of CBCT segmentation, IOS segmentation, and multimodal fusion. The deep learning based CBCT segmentation module is composed of domain-specific data preprocessing strategies, a feature extraction backbone based on the Swin transformer, and a novel segmentation head with auxiliary segmentation heads for pixel representation calibration. This module, which is termed as TSTNet, takes a single CBCT slice as input and generates predictions for each pixel (Fig. 1A). Specifically, the slices are first cropped to predefined sizes based on prior knowledge from input statistics. Afterward, multiple data augmentation methods, which had been demonstrated to be highly effective for CBCT segmentation, were employed for preprocessing. The segmentation backbone had a hierarchical Swin transformer which could extract the local features that can help identify the boundary between the background and tooth-bone structures, and global features which can provide richer context information for robust classification [29]. In CBCT slices, the area of the background class was much larger than that of the tooth class, which led to severe class imbalance problems. Hence, we proposed an auxiliary FCN head and a novel pixel-level tooth error calibrated representation learning head (TEC) to complement the main UpperNet head. Together with a hybrid loss including an Assignable Weight Online Hard Example Mining (AWOHEM) cross entropy loss, a novel metric loss and a Lovasz-Softmax loss, the TSTNet could attain fine-grained segmentation across various tooth morphologies [25, 33].

The IOS segmentation module, termed as IOSNet, aimed to generate predictions for each face in the IOS meshes. It first transformed the input mesh as point clouds, which was subsequently processed with the modified DCNet architecture [32, 18]. The DCNet had some limitations in precise boundary segmentation or generalization to complicated morphologies, e.g., crowded teeth or hyperdontia. To cope with these issues, we proposed two novel loss functions. The first included a centroid loss which helped learning the coarse tooth shapes to avoid weird segmentations (recognize two teeth as one tooth or vice versa). The other included a boundary loss which could help produce accurate boundary predictions for complex samples, i.e., mesh faces between the tooth-tooth and tooth-gingiva boundaries. The segmentation results over point clouds were further mapped back to the original meshes with a k-nearest neighbor aggregation strategy, followed by a standard graph-cut based smoothing for post-processing.

The multimodal fusion module, a.k.a. MFM, integrated the segmented IOS meshes and CBCT scans to generate fused, accurate, and high-resolution meshes with crown-root-bone structures for real-world clinical applications. This module first reconstructed the 3D mesh for the tooth and the alveolar bone from CBCT based on the marching-cube reconstruction and HLO smoothing algorithms [24]. However, the biting positions of CBCT and IOS were different, leading to contacts between the upper and lower jaws. The biting position difference and the contacts remained challenging for fusing both CBCT and IOS meshes. To address this issue, we proposed a novel point curvature estimation method to separate the upper and lower jaw in CBCT, which could be scarcely achieved using regular methods for contacted tooth models. Moreover, a two-stage registration method and a point-level mesh fusion method were proposed for data fusion. After refined with Laplacian smoothing, the final output showed precise delineation of the crown, root, and bone structures, which could be used for outcome simulation and treatment planning in many dental applications.

**CBCT Segmentation Results.** We conducted a five-fold cross validation test, each with a hold-out set of CBCT from 50 patients, to evaluate the segmentation performance of TSTNet (Table 1). We chose several widely used segmentation networks as our baseline, such as UNet, UNet++, Deeplabv3, FCN, and the standard Swin Transformer [26-30]. Other baselines which focus on instance segmentation were not compared here, as the task definition was different, and their codes were not publicly available. TSTNet achieved an average Dice coefficient and IoU of 93.99% and 88.68%, respectively, significantly outperforming the baselines. The precision-recall curve also demonstrated the superiority of our TSTNet. A further individual-level investigation on 43 patients revealed that TSTNet was able to obtain a 4.62% IoU and 2.73% Dice coefficient performance gain over the state-of-art baseline, demonstrating its robust generalization ability across patients.

**IOS Segmentation Results.** We evaluated the performance of IOSNet over the same test set as its pioneering work, DC-Net [18]. The test set consisted of 200 meshes from 100 patients. The performance was also compared against classical point cloud segmentation networks as well as a strong baseline, MeshSegNet [31]. The IOSNet achieved an mIoU of 93.07% and 95.70% on the maxillary and mandible IOS scans, which were 1.77% and 3.52% higher than best-performing baselines. Given that DCNet was already able to generate clinically applicable results for most cases [cite JDR paper], such a large improvement could corroborate better performance for the IOSNet, which was subsequently demonstrated by the real-world clinical demonstration with our DDMA framework.

**CBCT and IOS Fusion Results.** We evaluated the performance of three consecutive intermediate steps in the fusion module. First, the effectiveness of the proposed curvature-based algorithm for separating upper and lower jaws was tested on 50 patients, where 31 of them were in a biting position of tight tooth contacts. With appropriate threshold values, our method successfully separated 94% of the upper and lower jaws, while traditional Gaussian curvature or DBSCAN algorithms only separated 42% or 24% of them. In addition, different threshold values could be utilized to correct the failed cases. Second, we evaluated the accuracy of registration between the reconstructed CBCT mesh and IOS mesh for data fusion. The proposed two-stage registration method achieved much better performance in terms of fitness, inlier-rmse and corresponding set size on a test set of 30 jaws, leading to much higher success registration rate (91%) than selected baselines (13.3% or 6.67%), see Supplementary Tables. Lastly, we systematically evaluated the multimodal fusion quality by computing the average symmetric surface distance (ASSD), Housdoff distance (HD) and Chamfer Distance (CD) metrics between the fused outputs and ground truth provided by human experts. The corresponding distances were as small as ~0.2mm. Given that the resolution of CBCT machines were between 0.1-0.5mm, such a small distance error, as low as one pixel, would be acceptable for many clinical applications, as demonstrated by the clinical applicability test.

**Visualization.** Some representative segmentation results were shown in Fig 2, while a more comprehensive visualization was attached in the Supplementary. Our TSTNet committed much less false positive and false negative mistakes than the baselines (Fig. 2a). For example, the baselines might wrongly recognize the background as erupted third molars (case 1), fail to recognize the hyperdontia (case 2), or fail to distinguish the ambiguous boundaries among tiny tooth slices (case 3). The TSTNet generated much better results for these complicated cases. Similar performance could also be found in the jaw segmentation results as in case 4 and 5. Two cases were displayed to illustrate the superiority of our IOSNet for segmenting 3D IOS data (Fig. 2b). Together with more randomly chosen cases in the Supplementary, the IOSNet could generate segmentations nearly identical to the ground truth, as corroborated by comparing it to the almost clinically applicable baseline DC-Net.

The representative 3D reconstruction and fusion results were shown (Fig 3. a,b). By comparing the reconstruction with the fusion outputs, two main advantages could be identified for multimodal fusion: 1) the IOS segmentation could naturally provide FDI tooth number information complementary to CBCT semantic segmentations. 2) The fused outputs were associated with high-resolution tooth crown data from IOS and accurate tooth root and alveolar bone with a joint smoothing step, which were essential for diagnosis and treatment planning in dental practices. Meanwhile, the fused results were almost identical to the ground truth, with regards to the root-bone relationships (Fig. 3a). Notably, the unerupted third molars were not used in the ground truth for real-world clinical applications. More fused results could be found in the Supplementary.

**Ablation Study.** We conducted an ablation study to evaluate the effectiveness of proposed TSTNet and IOSNet, respectively. As for TSTNet, the original baseline, i.e., Swin transformer, only achieved an IoU of 90.52%, on a development set of 43 patients. In contrast, our TSTNet achieved an IoU of 93.22%, while each novel component, i.e., domain specific data augmentation, the weighted loss function, and the pixel calibration head, significantly contributed to the enhanced performance. For the IOSNet, the modified DC-Net could already achieve quite good performance, while adding a single centroid loss or boundary loss might lead to inferior performance.

However, the combination of these designs led to further performance gain. This was mainly due to the centroid loss and boundary loss, which were supposed to work together to help identify the position and produce clear tooth-tooth and tooth-gingiva boundaries. A single loss might impose strong biases in segmentation results.

**Clinical Utility and Demonstration.** To demonstrate the clinical utility of our DDMA framework in dental practices, we established two workflows for real-world clinical treatment planning with AngelAlign Tech in orthodontics. The first is to simulate crown movement using IOSNet (Fig. 4a), the most used method in the current clear aligner treatment planning (cite). The second is to simulate the whole tooth movement using our DDMA fusion framework (Fig. 4b), a transformative way to visualize crown-root-bone structures and relationships across the entire orthodontic treatment process.

We evaluated the end-to-end inference time of each module and the DDMA framework. The TSTNet took about 0.052 second (raw algorithm) and about 1 second with multi-scale post-processing to segment one CBCT slice. This was 3 to 4 orders of magnitude faster than human experts using the interactive Lableme software. The IOSNet was also about 50 times faster than human experts. The current DDMA framework took about 20 to 25 minutes to generate the fused 3D mesh model following the sequential processing order, which took at least 5 hours by experienced human experts, even with the help of interactive software and semi-automatic algorithms (Fig 1). The current inference speed is very appealing to dentists, as demonstrated by the integration of our method into real-world clinical software 'Intelligent Root System' (IRS) by AngelAlign Tech. However, in our current framework, CBCT slices were segmented one by one followed by the IOS segmentation before fusion. Such inference process could be substantially accelerated by parallelly processing the CBCT slices or segmenting the IOS and CBCT simultaneously.

To show the limitations of the IOSNet and the clinical significance of our DDMA framework, we selected two representative clinical cases using clear aligners. Figure 5 showed the patients were treated with IOSNet planning and showed insufficient tooth retraction after 10 months of treatment (Fig. 5a-d). No other problems were detected from intraoral examinations and the IOS treatment plan. By simulating the crown-root-bone structures using our DDMA, we found dehiscence in the canine tooth before the extraction (Fig. 5e), which might be the major cause of insufficient anterior tooth retraction. New treatment planning corrected the dehiscence before tooth extraction, which could accelerate tooth movement and avoid root resorption (Fig. 5g). Figure 6 showed the patient with the chief complaint to lower the anterior teeth with clear aligners (Fig. 6a). The patient was treated with IOSNet planning and showed insufficient tooth movement after 10 months of treatment (Fig. 6b-d). We simulated the crown-root-bone structures using our DDMA framework and identified multiple sites of fenestration at this stage (Fig. 6e), which might be the major cause of insufficient tooth movement. In addition, the previous IOSNet treatment planning will cause severe dehiscence after treatment, as identified by DDMA (Fig. 6f). New treatment planning corrected all fenestrations before lowering the anterior teeth (Fig. 6g). The simulated treatment outcome showed no sites of dehiscence or fenestration (Fig. 6h). Therefore, in both cases, the DDMA framework was able to identify the problems caused by root-bone relationships during orthodontic treatment, which could not be detected by human experts using IOSNet or intraoral examinations.

**Discussion**

The main contribution of our work is the development of the first deep learning based multimodal fusion framework for automatic and accurate tooth crown-root-bone reconstruction and analyses. From the clinical perspective, the value discrepancy and limitations of CBCT impede its clinical applications. The IOS scans, on the other hand, show the highest accuracy, but miss the information of tooth root and bone structures. The limitations in both CBCT and IOS models motivate the design of a multimodal fusion system. Notably, we demonstrate the effectiveness and clinical applicability of our DDMA framework in orthodontic treatment planning and risk prediction. From the technical perspective, DDMA is the first multimodal fusion framework which automatically segments CBCT and IOS data with novel deep neural networks. In addition, the two modalities are integrated with a novel fusion algorithm and our framework shows superior performance for CBCT and IOS segmentation with state-of-art accuracy. Furthermore, the biting positions for CBCT and IOS are different for one patient, making the fusion process challenging. Our DDMA framework can tackle the fusion problem with registration and fusion strategy with one jaw, where a novel curvature is defined to segment the upper and lower jaws. Therefore, our framework can work on a single jaw to avoid the different jaw positions of CBCT and IOS models.

The automatic segmentation of CBCT has been established by many groups, while achieving clinical applicable segmentation and tooth identification remains challenging. Some prior work formulated tooth segmentation in CBCT as an instance segmentation task on 3D CBCT images, which usually requires annotating both tooth pixels and tooth instances on 3D voxels across the entire CBCT scan [9]. In contrast, we recognize the 3D tooth and alveolar bone with semantic segmentation over 2D slices. Our TSTNet only requires annotating tooth pixels on several 2D slices, i.e., about 20 slices in a CBCT scan out of 300 to 600 slices, which can be transformed to instance annotations on-the-fly. Meanwhile, it is also critical to generate FDI tooth codes for clinical applications. Cui et al only generate tooth codes for cases with 28 teeth, but that cannot be applied to other tooth codes, e.g., third molars. Tae et al designed a tooth identification method based on 2D panoramic images reconstructed from CBCT to identify incisors, canines, premolar, and molars, followed by assigning FDI tooth codes accordingly [6]. However, the two-stage classification procedure is error-prone, and it cannot handle complicated cases like missing teeth or hyperdontia, as the tooth numbers are not inherently provided. In DDMA, we adopt the semantic segmentation strategy, while leaving the identification of the FDI tooth codes in the fusion step. By doing so, we substantially reduce human labor for annotation, as well as network complexity and corresponding training efforts for segmentation. For example, hierarchical or multi-stage networks are not needed [6-7,9]. Moreover, the IOS models can give much more accurate and comprehensive estimates for FDI tooth codes. While Cui's work only focuses on tooth segmentation, we segment teeth and alveolar bones simultaneously [7]. In addition, TSTNet obtains superior segmentation performance with novel auxiliary segmentation heads, loss functions, and augmentations, leading to more clinically applicable 3D models from CBCT.

In addition to CBCT segmentation accuracy, the quality of fused output also relies on the IOSNet and MFM modules. As for 3D tooth segmentation on IOS, the previous DC-Net can generate clinically applicable results for real-world clinical cases. However, our IOSNet surpasses its performance with a modified network, novel loss functions, and a larger dataset. Such a state-of-the-art performance corroborates clinically applicable segmentations for almost any case from the clinics. For the multimodal fusion of CBCT and IOS, previous work used traditional registration and level-set segmentation methods to fuse CBCT and IOS [22]. However, the level-set method requires interactive initial point selection, and the segmentation performance is inferior compared to deep neural networks. Besides, previous work relies on heuristic threshold values to crop the tooth crowns in CBCT, resulting in gaps in the fused meshes. In contrast, our DDMA framework proposes novel deep learning-based methods for both CBCT and IOS segmentation and achieves end-to-end fusion with a novel half jaw registration and fusion strategy. Notably, our curvature-based method shows a strong capability to segment upper and lower jaws, while traditional methods, like Gaussian curvature, failed in most cases. Automatically delineating the upper and lower jaws is an indispensable step for half-jaw registration and fusion. Consequently, DDMA can obtain more accurate segmentations and automatically generate fused meshes without manual intervention, as demonstrated by the comprehensive validation experiments with our large-scale datasets.

One major novelty of our work is the demonstration of clinical applicability. In collaboration with Angel Align, a clear aligner manufacturer, we developed the Intelligent Root System (IRS) software backed by our DDMA framework (Fig. 4b). Our model takes about 20 minutes to generate the fused 3D mesh model followed by a 5 to 10 minutes refinement, which is at least 12 to 15 times faster than human labors. Notably, the comparison was conducted when the current model had not been optimized with parallel processing, and human experts were allowed to use interactive software and semi-automatic algorithms. The IRS system has been used in real-world clinics to help doctors visualize the root-bone relationships during the entire orthodontic treatment and make better decisions for treatment planning. In addition, our real-world clinical cases proved the superiority of our system when dealing with problems of root-bone relationships, compared to traditional treatment planning using IOS. Therefore, we have demonstrated that our proposed deep learning approach can accurately recapitulate tooth crown-root-bone structures with an accuracy comparable to human experts. In addition, we also validated the clinical applicability of our model by saving 80% to 90% of a human expert's time spent on the crown-root-bone reconstruction. Furthermore, we have shown evidence with real-world clinical cases that our system can visualize root-bone structures and predict problems caused by root-bone relationships during orthodontic treatment. Our work could help clinical dentists make better treatment plans and avoid problems like dehiscence and fenestrations.

Although our DDMA framework has demonstrated its performance and clinical applicability, there are nevertheless some limitations for future work. First, the MFM module requires several intermediate steps such as curvature-based segmentation, two-stage registration, and point cloud-based 3D fusion. This brings the most mistakes in the DDMA framework, i.e., both the jaw segmentation and registration method might fail for 6% to 10% cases based on our statistics, implying manual correction is still required for those complicated cases. We might design novel learning-based methods in our future work for better jaw segmentation, 3D registration and fusion, such as taking advantage of the z-values of CBCT slices for jaw segmentation or design novel neural networks for 3D registration. Second, the segmentation performance for CBCT can be further improved. Though TSTNet achieves superior performance even for patients with unerupted teeth, hyperdontia, malposition, or ambiguous tooth boundaries, it might still make mistakes for complicated cases, e.g., failing to identify some metal artifacts or patients with root canal therapy. Future work might combine deep learning with domain knowledge in stomatology for better segmentation in both CBCT and IOS. Finally, though our method is already integrated into the clinical software to assist orthodontic treatment planning, the current evaluations are more from the algorithmic perspective, while clinical usage should be further demonstrated with rigorous multi-center clinical trials for large-scale various complicated cases.

**Conclusion**

In this paper, we propose a novel DDMA framework of multimodal CBCT and IOS fusion for intelligent tooth crown-root-bone analysis, consisting of a CBCT segmentation module, an IOS segmentation module and a multimodal fusion module. The effectiveness of each individual module as well as the entire framework is systematically demonstrated with comprehensive experiments on our large-scale multimodal dataset. Furthermore, the clinical utility and applicability is reported and demonstrated with real-world cases. Our framework has been integrated into clinical software to assist dentists in orthodontics treatment planning with a short demo in the Supplementary. Future work includes better multimodal fusion and segmentation algorithms, as well as large-scale multi-center clinical trials.

**Methods**

**DDMA Overview.** The DDMA framework consists of three major modules: TSTNet for CBCT segmentation, IOSNet for IOS segmentation and MFM for multimodal fusion, respectively, as illustrated in Fig 1. Below we would like to elaborate the details.

**1. CBCT Segmentation with TSTNet**

Given the CBCT slices, the first step of our method is performing 2D segmentation on each CBCT image. The goal of this step is to obtain a pixel-wise tooth mask and alveolar bone masks, which can be used to reconstruct the 3D tooth and alveolar bone mesh models in the following steps.

**Data Preprocessing.** The number of background and tooth pixels in the CBCT images are highly imbalanced. Based on empirical statistics from the annotated masks, we cropped the lower 1/4 and right 1/10 in the original CBCT image to partially alleviate the class imbalance problem. Afterwards, the CBCT images are resized to random resolution within 2048*2048, and subsequently randomly clipped to 512*515 and flipped. In the test phase, a multi-scale strategy is introduced to integrate multi-scale information for improved segmentation.

**Deep learning-based segmentation.** To perform 2D segmentation on each CBCT slice, a deep learning method based on Swin Transformer is proposed, named Tooth Swin Transformer Network, a.k.a. TSTNet. As shown in Supplementary information, TSTNet is composed of a backbone network and multiple segmentation heads. Swin Transformer is utilized as the backbone, which is a hierarchical Transformer whose representation is computed with Shifted windows. The hierarchical architecture has the flexibility to model at various scales, which can extract both local features and global features. The local features help to identify the boundary between the background class and the tooth class, while the global features provide richer context information for robust classification. The shifted windows bring the capability of modeling long-range dependency while maintaining the efficiency, which improves the classification precision further. Such designs make the Swin Transformer achieve superior performance on multiple vision tasks compared to CNN based methods, which is suitable for the backbone of TSTNet. Concretely, the backbone of TSTNet contains 4 stages. Each stage is built with 2/2/18/2 swin transformer blocks separately, of which the multi-head attention module (MHA) contains 4/18/16/3 heads respectively. The embedding dimension of each block is set to 128. See architecture details in the Supplementary.

We propose to use a main UpperNet segmentation head and two auxiliary segmentation heads, i.e., the FCN head and the tooth error calibration head (TEC), for segmentation. The UpperNet head is used for multiscale feature aggregation and fusion. Concretely, it is composed of a Pyramid Pooling Module (PPM) and a Feature Pyramid Network (FPN). PPM takes the last layer of the backbone as input and extracts global features at four pyramid scales with adaptive average pooling, which significantly enlarges the receptive field of view, bringing effective global prior representations for tooth or alveolar bone pixel classification. FPN fuses feature maps at different scales to generate a high-resolution feature map with both local features and global features, making the predicted tooth or alveolar bone mask to have a precise boundary[29].

In CBCT slices, the area of the background class is 10 times larger than that of the tooth class, which leads to the severe class imbalance problem. To overcome this issue, Assignable Weight Online Hard Example Mining (AWOHEM) cross entropy loss is introduced as the objective function of the UperNet head, which avoids overfitting of the background class. The AWOHEM loss tends to choose the examples with higher loss or more diversity as the training data and assign the different weights of classes, reducing the bias to the class with majority samples [33].

We propose the TEC can calibrate such error-prone feature representations in boundary areas for better segmentation performance. Notably, TEC could be incorporated into the hidden layers, trained together with the whole TSTNet, and decoupled in the inference stage without additional parameters and inference time. The TEC head consists of the image decoupling, prototype clustering, and error calibration modules. The ground truth and the predicted tooth mask of the UperNet head server as inputs for TEC. The image decoupling module categorizes the pixels of the prediction into 3 sets: True Positive (TP), False Negative (FN), and False Positive (FP) $s_k^{tp}, s_k^{fn}, s_k^{fp}$ for category k. Based on current and historical TP pixels, the prototype clustering module calculate the category prototype via EMA ((Ye et al. 2019; Wu et al. 2018)):

$$\mu_k = \rho\mu_k + (1-\rho)\frac{1}{n_k^{tp}}\sum_{i \in s_k^{tp}} \mathbf{e}_i$$

where $n_k^{tp}$ is the number of current TP pixels for category k, $\mathbf{e}_i$ is the embedding for pixel *i*, $\rho$ is the momentum value to adjust the retained proportion of historical prototype. Afterward, the cosine similarity between pixel i and prototype $\mu_k$ in embedding space is defined as follow:

$$\cos\theta_{ik} = \tilde{\mu}_k \tilde{e}_i^\top = \frac{\mu_k e_i^\top}{\|\mu_k\|_2 \|e_i\|_2}$$

where $\|\cdot\|_2$ is the L2 distance and $\tilde{\mu}_k$ is the normalized vector with magnitude 1. Our method aims at maximizing $\cos\theta_{ik}$ to make $e_i$ close to prototype $\mu_k$, thus clustering pixels of the same category to better tooth segmentation performance.

TEC parallelly calculates the FN and FP penalty terms for error segmentation pixels. The FN penalty term pulls the FN pixels towards the anchor of the tooth class, while the FP penalty term pushes the FP pixels to the opposite pole against the anchor in the embedding space. The FP errors occur when pixels of other categories are excessively similar to category k in the feature space. On the other hand, the FN errors occur when pixels belonging to category k are misclassified as other categories. To tackle this, we take pixel i of category k as an anchor. We defined the FN and FP penalty terms as follow:

$$T_i^{fp} = \begin{cases} 1 + \frac{1}{n_k^{fp}} \sum_{j \in s_k^{fp}} \tilde{\mathbf{e}}_j \tilde{\mathbf{e}}_i^\top &, \text{if } n_k^{fp} > 0 \\ 0 &, \text{otherwise} \end{cases}$$

$$T_i^{fn} = \begin{cases} 1 - \frac{1}{n_k^{fn}} \sum_{j \in s_k^{fn}} \widetilde{\mathbf{e}}_j \widetilde{\mathbf{e}}_i^\top & , \text{ if } n_k^{fn} > 0 \\ 0 & , \text{ otherwise} \end{cases}$$

$T_i^{fp}$ is 0 when there are no FP pixels or the average similarity converges to -1, indicating all FP pixels are in the opposite direction from anchor i. $T_i^{fn}$ is reduced to 0 when there are no FN pixels, or all FN and TP pixels are placed at the same direction in the feature space. Finally, TEC integrates the cosine similarities and the FP/FN penalty terms into a TEC loss, serving as the objective function to train the whole network.

$$L_i^{TEC} = -\log \frac{e^{\cos\theta_{ik}/\tau - (1-p_{ik})T_i^{fn}}}{e^{\cos\theta_{ik}/\tau - (1-p_{ik})T_i^{fn}} + \sum_{l \neq k} e^{\cos\theta_{il}/\tau}} - \log \frac{e^{\cos\theta_{ik}/\tau - (1-p_{ik})T_i^{fp}}}{e^{\cos\theta_{ik}/\tau - (1-p_{ik})T_i^{fp}} + \sum_{l \neq k} e^{\cos\theta_{il}/\tau}}$$

where $\tau$ is the temperature hyper parameter. $p_{ik}$ comes from the error segmentation pixel. Overall, the total loss for Upernet head is defined as:

$$L_i^{seg} = L_i^{\text{AWOHEM}} + \lambda L_i^{TEC}$$

where $\lambda = 0.1$ is the factor to adjust the strength of the two losses.

Besides the UperNet head, TSTNet also employs an FCN head, serving as the auxiliary head to improve the segmentation performance further. As for the FCN head, Lovasz-Softmax loss is employed as the objective function. Lovasz-Softmax loss is better at segmenting small objects and reducing false negatives, which can solve the class imbalance problem to a degree and avoid the missed detection of tooth area [25].

### 2. IOS segmentation with IOSNet

The goal of the deep learning-based segmentation system is to perform point-wise classification for the point cloud obtained in the data preprocessing step. The data preprocessing follows the DC-Net baseline to convert the original IOS mesh to a point cloud with 15-dimensional features as in the DC-Net [18].

**Deep learning segmentation.** The architecture of the deep learning-based segmentation system is illustrated in Supplementary Figure x, which is modified from the Dynamic Graph CNN (DGCNN) and DC-Net. The main contribution in IOSNet is the proposal of two additional loss functions: the centroid loss and the boundary smoothing loss. More specifically, given the tooth point cloud $P = \{p_i\}, i = 1 \text{ to } 10{,}000$ with 10,000 points, the centroid loss is defined as :

$$L^{centroid} = \frac{1}{C} \sum_{i=1}^{C} dis(pc_i - gc_i)$$

where $C$ denotes the number of categories in the annotation, $dis(\cdot)$ denotes the Euclidean distance of two points, $pc_i$ and $gc_i$ denotes the prediction centroid and the

gold centroid for the $i$-th class, respectively, which are concretely defined as:

$$pc_i = \frac{\sum_{j=1}^{N_i} \widetilde{p}_j \times s_j}{\sum_{j=1}^{N_i} \widetilde{p}_j}$$

$$\widetilde{p}_j = \begin{cases} 1, & p_j > th_u, \\ 0, & p_j < th_l, \\ p_j, & otherwise. \end{cases}$$

$$gc_i = \frac{\sum_{j=1}^{M_i} s_j}{M_i}$$

where $N_i$ and $M_i$ denote the number of points in the $i$-th class except the gingiva in the prediction and the ground truth, $p_j$ and $\widetilde{p}_j$ denote the probability of the $j$-th point predicted as the $i$-th class before and after processing, and $s_j$ represents the 3D coordinates of the $j$-th point. Here, we define $th_l$ and $th_u$ as 0.38 and 0.6, respectively.

The boundary loss is defined as:

$$L^{boundary} = L^{ce}(pb, gb)$$

where $L^{ce}$ denotes the cross-entropy loss, $pb$ and $gb$ represent the prediction probability and ground truth label of the boundary points, respectively. Here, the top 5% points are chosen as the boundary points according to the *KL-divergence* of all the points, while the KL-divergence for the $i$-th point is defined as

$$KL\_div_i = \max_{j \epsilon \kappa_i} KLD(c_i, k_j)$$

where $c_i$ means the center point's prediction probability distribution, and $k_j$ denotes the $j$-th neighbor point's probability distribution, KLD is the KL-divergence between two distributions. We use maximum as the aggregation operation in a local neighborhood to select the most representative value from a set of KLD values computed by the center point and its each neighbor point as the center point's *KL_div*. It indicates the distributional difference between the $i$-th point and its $K$ surrounding points, which can be used to measure the possibility that this point becomes a boundary point. We set K=5 in our implementation. The corresponding normalization and activation layers are changed for better performance. Moreover, we use the graph-cut based boundary smoothing component to further improve the segmentation performance as in the DC-Net [18]. Architecture details were shown in the

Supplementary information.

## 3. Multimodal Data Fusion

The MFM module segments the jaws in the reconstructed CBCT mesh with a novel curvature-based segmentation algorithm, and further replaces the unsatisfactory crown with high-quality laser-scanned crown mesh. The MFM modules includes three steps: (1) 3D CBCT reconstruction and automatic segmentation of half jaws and individual teeth in the reconstructed mesh based on point curvature; (2) Registration of CBCT data and crown mesh (maxillary and mandibular teeth, respectively); (3) Replacement of the crown of the reconstruction model to obtain an accurate result by a fusion algorithm.

**CBCT Mesh Reconstruction.** We employ the marching cubes algorithm to obtain the complete reconstructed CBCT mesh of teeth and jaws. However, the result is always unsatisfactory, especially its crown part, due to three main problems: low resolution, tooth crowding, and the occlusal problem. The HLO (Pan et al. 2020) is applied to smooth the surfaces. Compared with other filters, HLO can maintain the boundary well while reducing the sharp noises, with its edge-preserving features.

**Point Curvature based Segmentation.** The reconstructed CBCT meshes usually have connected boundaries between adjacent teeth or the contacts in maxillary and mandible, especially for patients in a close bite position, imposing great difficulties for accurate half jaw registration and individual teeth delineation. Hence, we propose a new segmentation algorithm to separate the half jaws and each tooth using a novel point curvature. Specifically, the point curvature of a vertex is defined as the average of angles between normal vectors of all its neighbors, which is different from either mean curvature or Gaussian curvature. Mathematically, given an vertex $v$ with normal vector $n_v$, we defines its first-order neighbors as $u \in N(v)$, associated with normal vectors $n_u$, the point curvature $c_v$ of vertex $v$ is defined as:

$$c_v = \frac{1}{|N(v)|} \sum_{u \in N(v)} arccos(\frac{n_v \cdot n_u}{|n_v||n_u|})$$

This definition can be extended to $l$-th-order neighbors as well by changing the set of neighbors. The intuition behind the curvature-based segmentation algorithm is that the angle between the normal vectors of adjacent points can capture the curvature of different scales by changing the level of neighbors. We empirically found that this curvature can clearly recognize the ambiguous boundaries between adjacent teeth, and the upper and lower jaw contacts.

The point curvature segmentation algorithm is illustrated in Supplementary Algorithm 1-3. The algorithm is like an "erosion-expansion" procedure. We first compute the point curvature for each vertex, and vertices with curvature in the top $M$ percent would be removed from the mesh (erosion). Followed by a simple connected component analysis algorithm, the individual teeth can be separated. The deleted vertices would be merged back to the nearest separated components, on top of which the KD-trees are constructed, to get unbroken reconstructed teeth (expansion). Based on each individual tooth, we can easily get the mandible and maxillary by computing the gravity center of each tooth and separating them into upper and lower parts with a simple ransac algorithm.

**Registration of two modalities.** The segmented CBCT results from the previous section are provided as mandible and maxilla parts separately, which are registered with the IOS meshes. We denote the vertex set of the IOS mesh as $V_I$, and the vertex set of the CBCT mesh as $V_C$. The registration is two-stage, i.e., a global initial alignment is proposed to roughly register the two-point sets as the initial alignment of the two-point sets are of great difference, and an improved ICP registration step for further refinement. The global registration contains two steps: (1) Scale alignment: The scale of IOS mesh is adjusted by the pixel spacing and slice thickness properties of CBCT data, which is provided by the CBCT scanning machine. (2) Global rigid transformation: The CBCT and IOS meshes are first down sampled and then registered with a RANSAC scheme on the 33-dimensional FPFH vector space. Afterwards, the global registration results are further improved by the multi-scale ICP algorithm performing point-to-plane ICP registration at three scales.

**Tooth crown replacement by mesh fusion.** After registration of the CBCT and IOS, we propose a fusion algorithm to fuse the two modalities. The fusion is performed on the point cloud level. Specifically, we take the IOS point cloud as reference, removing the points belonging to the CBCT crown. We identify these points based on the Euclidean distance of the points in CBCT to the KDTree constructed from IOS point clouds. The points are sorted based on this Euclidean distance, and a heuristic portion of these points are removed. There might be points that belong to the CBCT crown but not removed. These points could be deleted by applying a simple DBSCAN algorithm on the rest points in CBCT that detect such isolated points. Finally, we can get the reconstructed mesh based on the fused point cloud and the original normals from the IOS and reconstructed CBCT meshes. The details of the fusion process could be found in Supplementary information.

**Statistical Analysis.** The mathematical formulation for segmentation metrics, such as IoU, Dice, ASSD, CD are defined in the main paper and Supplementary. All statistical analysis is conducted with Python and the corresponding packages such as sklearn and scipy.

**Reproducibility.** The model and its performance are verified with independent test cohorts. We also verified the performance of DDMA with multiple random seeds or initializations on different hardware platforms.

## Data Availability
The clinical CBCT and IOS data were collected by the hospitals in de-identified format. Owing to patient-privacy constraints, they are not publicly available. All requests for academic use of raw and analyzed data can be sent to the corresponding authors. All requests will be reviewed within 7-10 working days to determine whether the request is subject to any patient-confidentiality or intellectual property obligations. All requests will be processed in concordance with institutional and departmental guidelines and will require a material transfer agreement.

## Code Availability
The implementation code could be shared for academic use upon request.

## Acknowledgements
We acknowledge helpful discussions with C.X.Liu, Z.Chen, H.H.Yang, G.A.Wang, H.L.Wang.

## Competing interests
Y.F. is employed with AngelAlign Tech.

## Figure legends

**Figure 1. Overall pipeline of the DDMA framework that consists of three steps. A.** The pipeline of CBCT image segmentation for tooth and alveolar bone with TSTNet. **B.** The pipeline of IOS segmentation with IOSNet and post-processing to generate FDI tooth codes. **C.** The pipeline of the multimodal fusion module that fuses the reconstructed CBCT 3D meshes and IOS meshes with novel curvature-based segmentation, registration, and fusion methods.

**Figure 2. Visualization of the segmentation results for CBCT images and IOS meshes. A.** Five cases for tooth and alveolar bone segmentation, demonstrating that our method commits much less mistakes for false positives and false negatives. **B.** Two cases for IOS tooth segmentation, demonstrating that our method performs much better than the state-of-the-art baselines.

**Figure 3. Visualization of the reconstruction, fusion, and statistical results. A.** Reconstructed 3D CBCT meshes with marching cube and HLO algorithms. **B.** Two cases of fused 3D tooth and alveolar bone of DDMA and ground truth. **C.** Statistical results of the TSTNet segmentation results.

**Figure 4. Overall pipeline of orthodontic treatment planning and simulation with and without DDMA. a.** The workflow of orthodontic treatment planning and simulation using IOSNet segmentation. **b.** The workflow of orthodontic treatment planning and simulation using DDMA fusion framework to visualize crown-root-bone structures.

**Figure 5. One example of a clinical case using DDMA to predict the cause of delayed anterior tooth retaction with clear aligners. a.** The intraoral photo of the patient before orthodontic treatment. The chief complaint was to retract the anterior teeth. **b.** Treatment planning using IOSNet. The dark blue indicated the teeth before treatment, and the white simulated the outcome after treatment. **c.** The intraoral photo of the patient after 10 months of orthodontic treatment. **d.** Insufficient anterior teeth retraction was detected at this stage. No other problems were detected from intraoral examinations and the IOS treatment plan. **e.** Visualization of crown-root-bone structures using DDMA framework. Fenestration was detected in the anterior teeth, which was the major cause of delayed teeth retraction. **f.** The simulated outcome of the previous treatment using IOSNet planning. Tooth dehiscence and fenestration were detected in the simulated outcome. **g.** New treatment planning using DDMA to correct the position of anterior teeth to get rid of fenestration before retraction. **h.** The outcome simulation with the new treatment planning with DDMA framework.

**Figure 6. One example of a clinical case using DDMA to predict the cause of insufficient tooth movement with clear aligners. a.** The intraoral photo of the patient before orthodontic treatment. The chief complaint was to lower the maxillary anterior teeth. **b.** Treatment planning using IOSNet. The dark blue indicated the teeth before treatment, and the white simulated the outcome after treatment. **c.** The intraoral photo of the patient after 10 months of orthodontic treatment. **d.** Insufficient tooth movement was detected at this stage (tooth lowering is not effective). No other problems were detected from intraoral examinations and the IOS treatment plan. **e.** Visualization of crown-root-bone structures using DDMA framework. Fenestration was detected in the anterior teeth, which was the major cause of insufficient tooth movement (shown by red arrow). **f.** The simulated outcome of the previous treatment using IOSNet planning. Dehiscence and fenestration were detected in the simulated outcome. **g.** New treatment planning using DDMA to correct the position of anterior teeth to correct fenestration. **h.** The outcome simulation with the new treatment planning with DDMA framework. No dehiscence or fenestration was detected.

# Figure 1

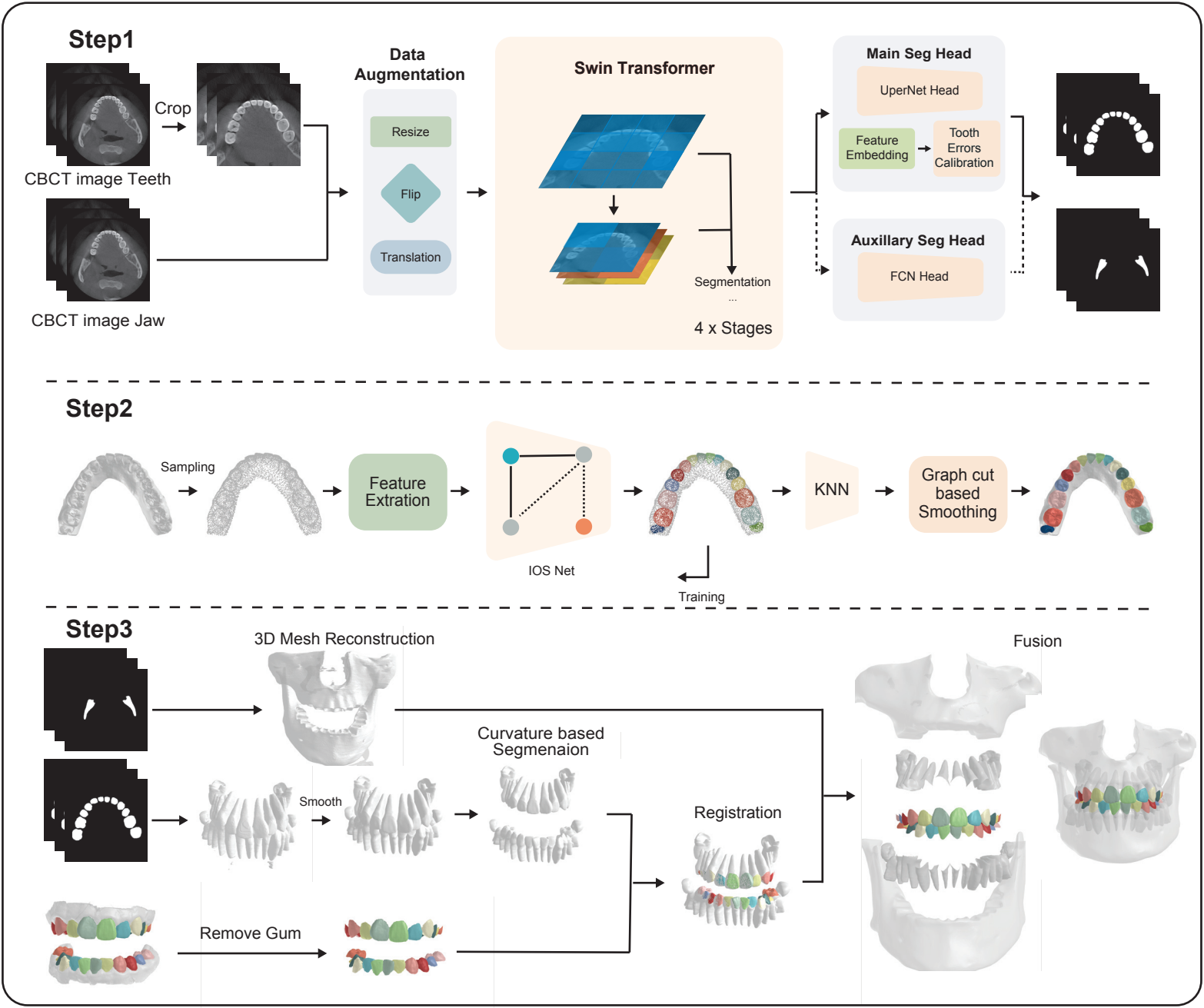

# Figure 2

**a** CBCT segmentation visualization

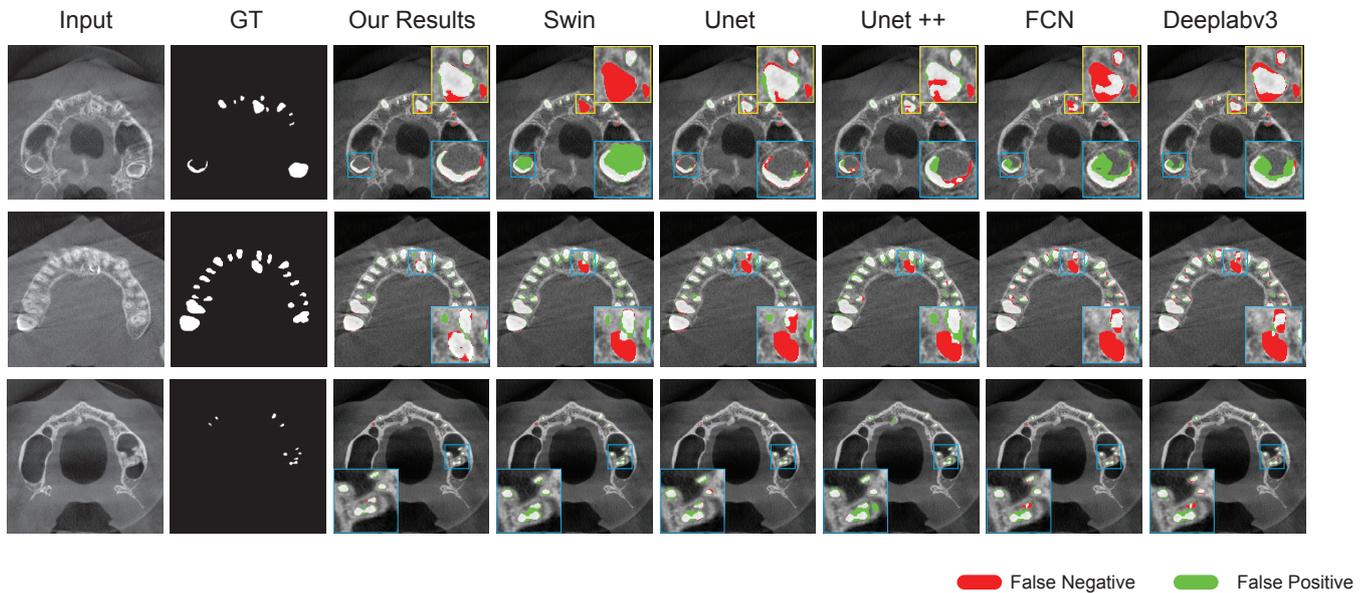

**b** IOS segmentation visualization

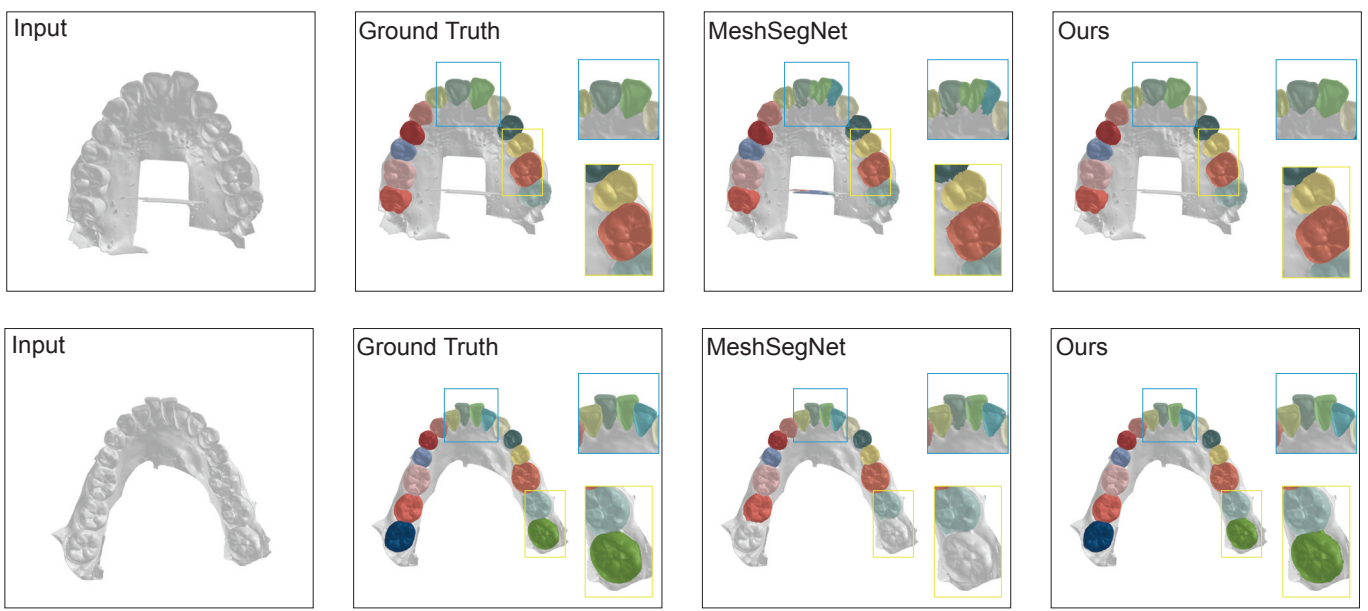

**Figure 3**

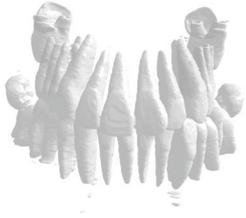
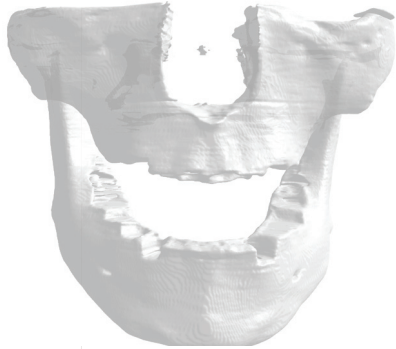
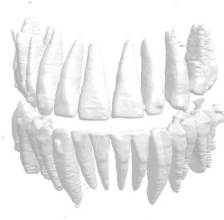
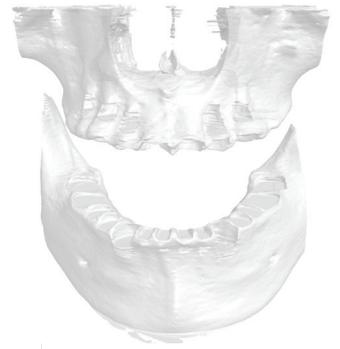
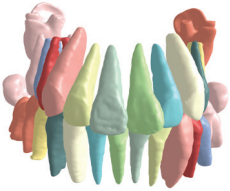
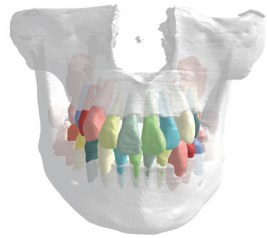
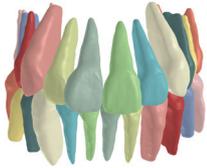
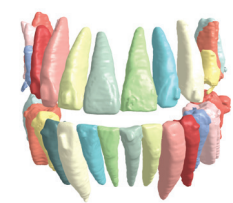
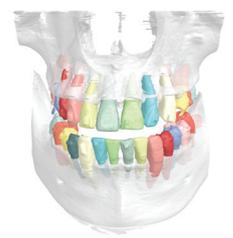
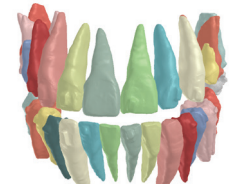
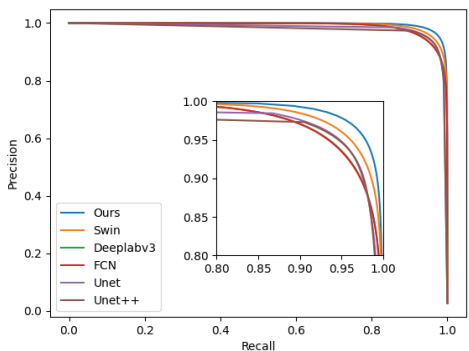
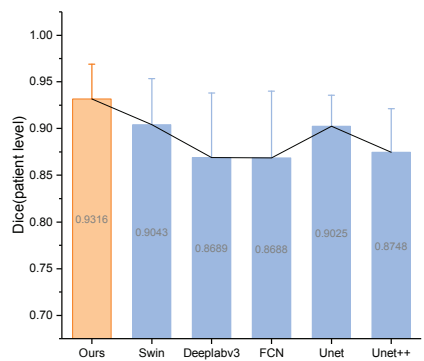
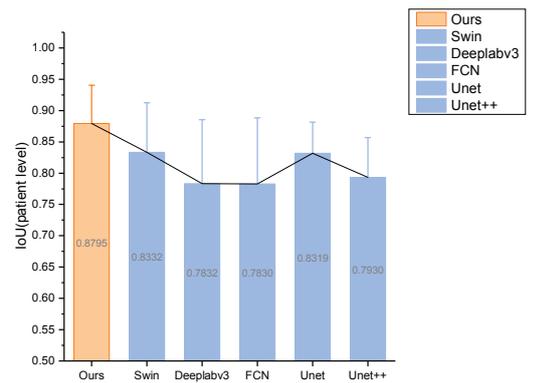

# Figure 4

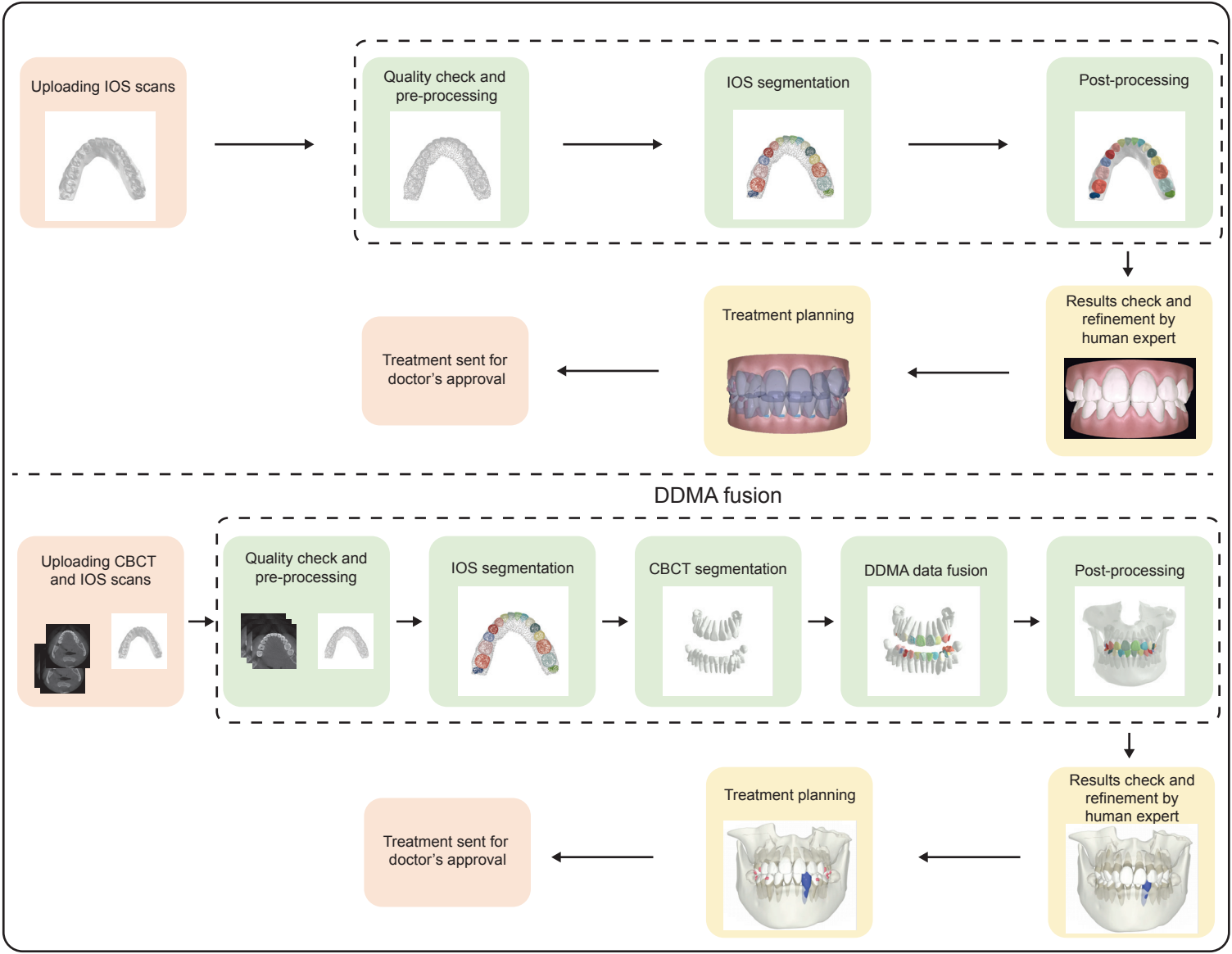

# Figure 5

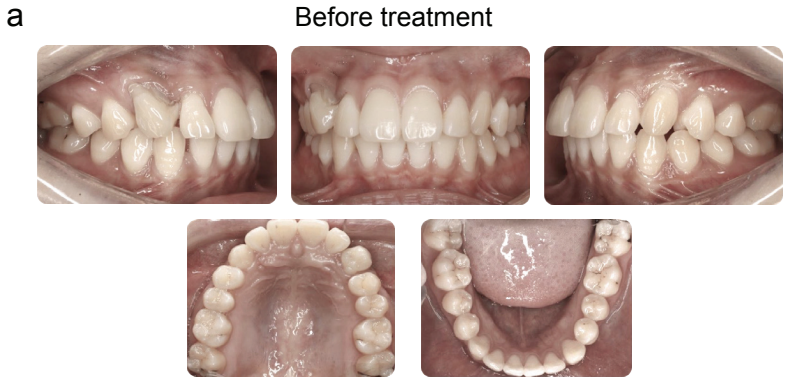
a. Before treatment

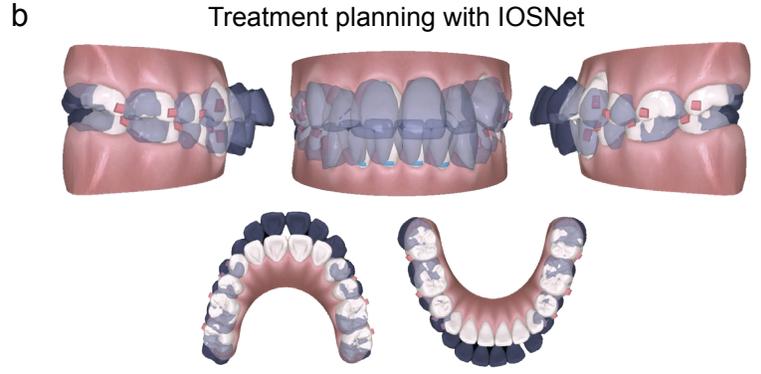
b. Treatment planning with IOSNet

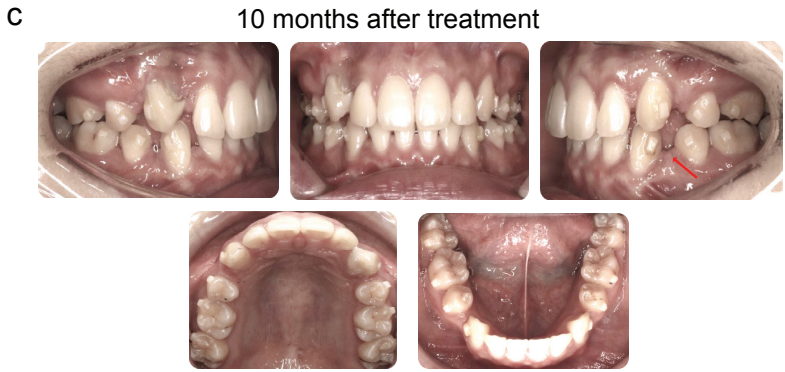
c. 10 months after treatment

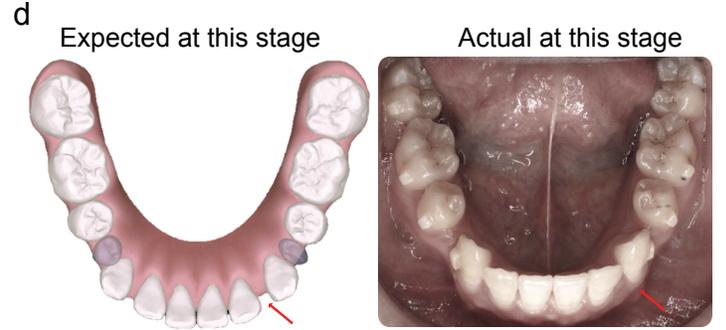
d. Expected at this stage / Actual at this stage

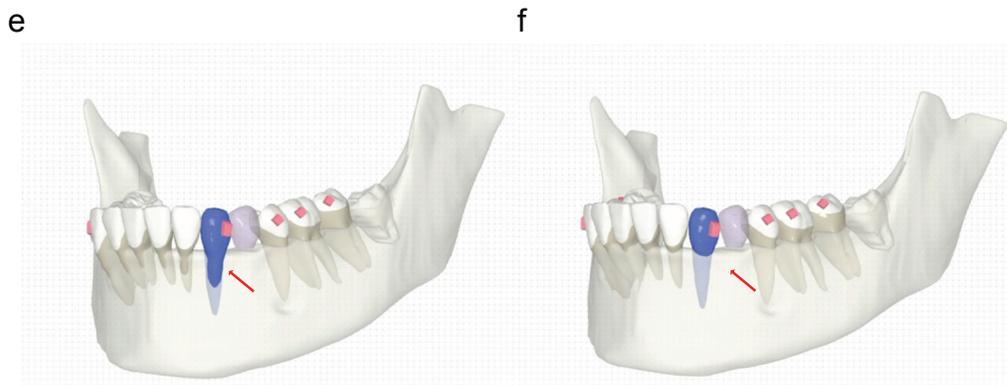

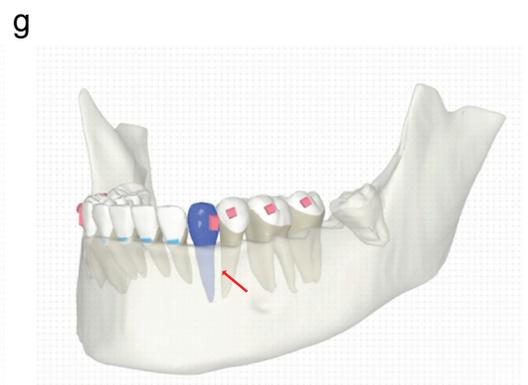

e. Visualization of crown-root-bone identified dehiscence

f. New treatment planning using DDMA to correct dehiscence

g. New treatment outcome using DDMA

# Figure 6

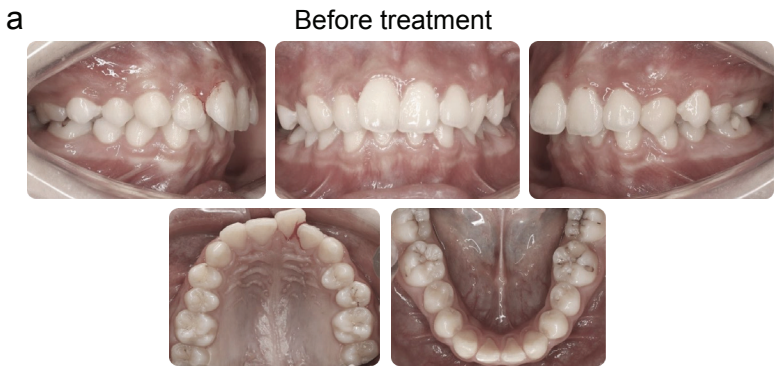

a  Before treatment

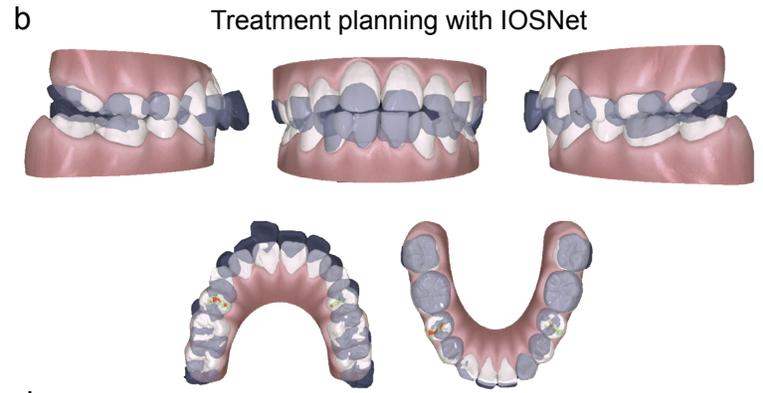

b  Treatment planning with IOSNet

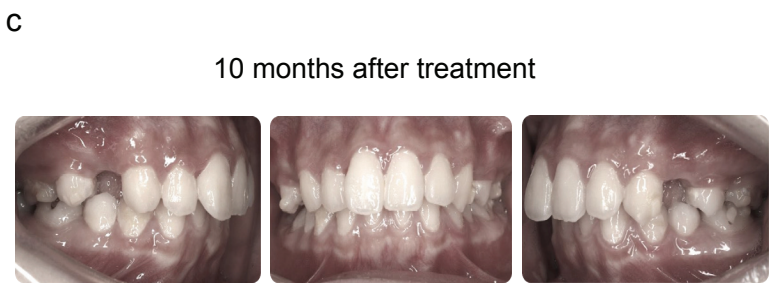

c  10 months after treatment

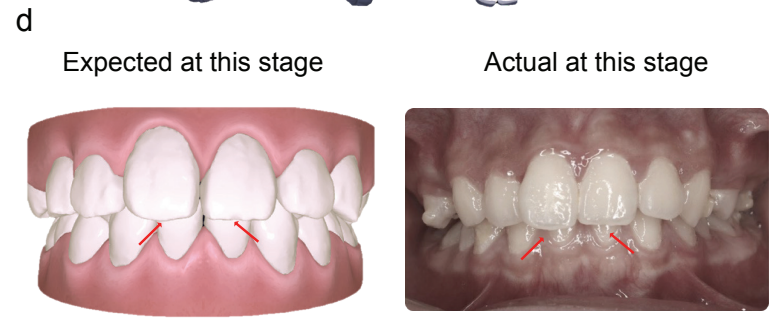

d  Expected at this stage  Actual at this stage

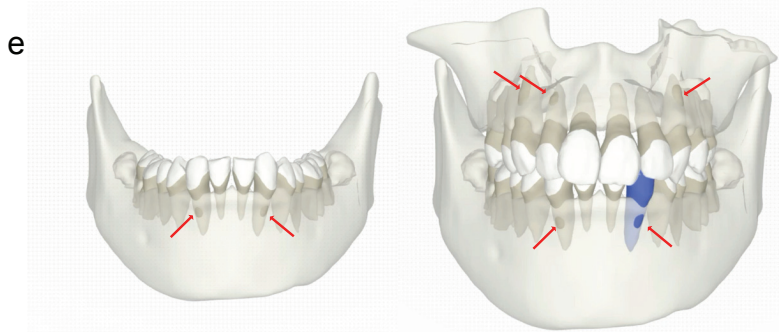

e  Visualization of crown-root-bone identified fenestration

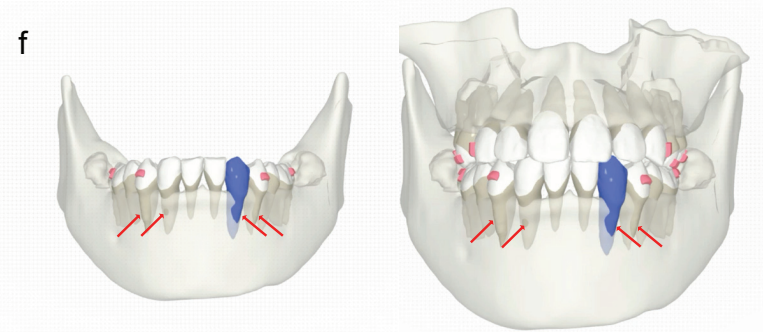

f  Treatment outcome simulation with IOSNet planning

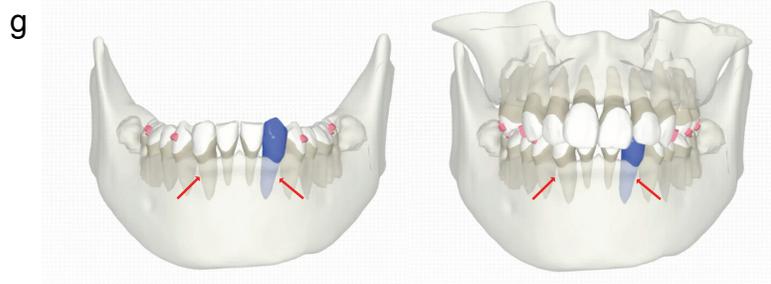

g  New treatment planning using DDMA to avoid fenestration

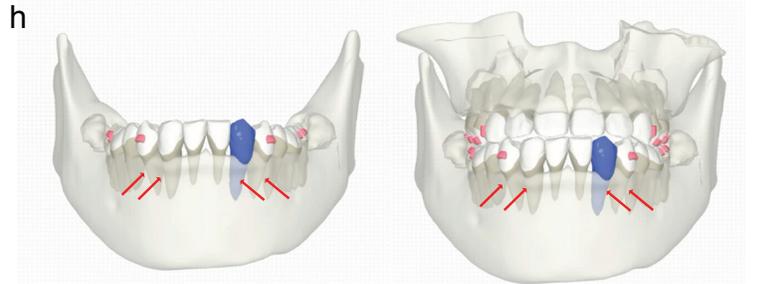

h  Treatment outcome simulation with DDMA planning

| Table 1. Five-fold cross-validation segmentation results on CBCT (each tested with n=50 patients). | | | | |
|---|---|---|---|---|
| **Model** | **Dice** | **IoU** | **Recall** | **Precision** |
| UNet | 92.95±0.61 | 86.90±1.06 | 92.83±2.74 | 93.23±1.57 |
| UNet++ | 92.95±0.66 | 86.83±1.16 | 93.40±1.66 | 92.51±0.60 |
| FCN | 92.28±0.92 | 85.68±1.60 | 92.21±2.31 | 92.39±0.48 |
| Deeplabv3 | 92.28±0.93 | 85.68±1.61 | 92.05±2.34 | 92.54±0.53 |
| Swin[3] | 93.29±0.67 | 87.61±1.19 | **93.95±1.68** | 92.95±0.54 |
| **TSTNet** | **93.99±1.20** | **88.68±2.15** | 92.37±2.78 | **95.71±0.70** |

IoU: intersection over union ; Dice: Dice Coefficient. Bolden numbers indicate the best performance.

The numbers denote the mean and standard deviation on cross validation.

| Model | Mandible | | | Maxillary | | |
|---|---|---|---|---|---|---|
| | mIoU(%) | $ACC_f$(%) | $ACC_a$(%) | mIoU(%) | $ACC_f$(%) | $ACC_a$(%) |
| CNN | 85.32 [81.96, 88.69] | 91.74 [89.77, 93.72] | 93.75 [92.14, 95.36] | 89.68 [86.90, 92.48] | 94.14 [92.35, 95.93] | 95.80 [94.37, 97.22] |
| PointNet | 63.10 [57.90, 69.29] | 79.67 [76.32, 83.01] | 86.21 [83.93, 88.49] | 59.05 [53.14, 64.95] | 75.58 [71.63, 79.53] | 80.95 [77.67, 84.24] |
| PoiintNet++ | 83.22 [81.38, 85.06] | 91.49 [90.41, 92.57] | 94.59 [93.91, 95.27] | 85.82 [84.14, 87.50] | 93.15 [91.92, 94.38] | 95.65 [94.95, 96.35] |
| DGCNN | 84.93 [83.27, 86.58] | 92.74 [91.82, 93.65] | 95.80 [95.18, 96.43] | 88.70 [87.55, 89.86] | 94.41 [93.73, 95.10] | 96.94 [96.42, 97.45] |
| MeshSegNet | 82.82[80.51, 85.13] | 93.50[92.55, 94.45] | 95.04[94.18, 95.90] | 85.62[83.61, 87.64] | 93.96[93.04, 94.88] | 95.59[94.77, 96.41] |
| DCNet | 91.93 [91.09, 92.78] | 96.01 [95.42, 96.61] | 97.98 [97.56, 98.41] | 92.18 [91.02, 93.36] | 95.99 [95.28, 96.71] | 97.90 [97.33, 98.45] |
| IOSNet | 90.37[88.69, 92.06] | 96.82[96.44, 97.20] | 97.96[97.61, 98.31] | 92.50[91.44, 93.57] | 97.11[96.79, 97.44] | 98.27[97.97, 98.57] |
| IOSNet* | 92.44[91.50, 93.37] | 97.31[97.05, 97.57] | 98.41[98.18, 98.64] | 93.73[93.21, 94.25] | 97.36[97.10, 97.62] | 98.53[98.32, 98.74] |
| IOSNet** | **93.70[92.15, 95.25]** | **98.14[97.78, 98.50]** | **98.81[98.49, 99.14]** | **95.70[94.80, 96.60 ]** | **98.35[98.08, 98.61]** | **99.05[98.82, 99.28]** |

w/o: without. mIoU: mean Intersection over Union, $ACC_f$: per-face accuracy, $ACC_a$: average-area accuracy.

[] indicates 95% confidence intervals. Bolden numbers indicate the best performance.

All baselines and IOSNet are trained on a small dataset with 4271 IOS scans.

IOSNet*: IOSNet trained on 28,000 scans without smoothing. IOSNet**: IOSNet* with smoothing.

Table 3.1. Tooth multimodal data fusion Metrics of the proposed method. (tested on 1003 teeth in 50 cases).

| Metrics | ASSD | CD | HD |
|---|---|---|---|
| Error(mm) | 0.18±0.07 | 0.20±0.06 | 0.38±0.17 |

ASSD:Average symmetric surface distance; CD: Chamfer distance; HD: Housdoff distance;

Table 3.2. End-to-end Model inference time and Clinical Utility of DDMA.

| Models | DDMA | Human Experts |
|---|---|---|
| Inf-T(minute) | ~20-25 | ~300-400 |

Inf-T: end-to-end inference time. ~ means approximately.